\documentclass{emulateapj}

\usepackage{epsf}

\newcommand{\ie}{\emph{i.e.,} }
\newcommand{\eg}{\emph{e.g.,} }

\newcommand{\gama}{$\gamma$}
\newcommand{\lsd}{l_{sd}}
\newcommand{\omp}{\omega_p}
\newcommand{\hbx}{\mathbf{\hat x}}

\newcommand{\ave}[1]{\left\langle #1\right\rangle}

\begin{document}

\title{Magnetic field evolution in relativistic unmagnetized collisionless shocks}

\author{Uri Keshet\altaffilmark{1}\altaffilmark{4}\altaffilmark{5},
Boaz Katz\altaffilmark{2}, Anatoly Spitkovsky\altaffilmark{3}, Eli Waxman\altaffilmark{2}}

\altaffiltext{1}{Institute for Advanced Study, Einstein Drive, Princeton,
NJ, 08540, USA}

\altaffiltext{4}{Benoziyo Center for Astrophysics, Weizmann Institute,
Rehovot 7600, Israel}

\altaffiltext{5}{Department of Astrophysical Sciences, Princeton
University, Princeton, NJ 08544, USA}

\altaffiltext{4}{Current address: Center for Astrophysics, 60 Garden St., Cambridge, MA 02138, USA}

\altaffiltext{5}{Einstein fellow}

\date{\today}

\begin{abstract}
We study relativistic unmagnetized collisionless shocks using unprecedentedly
large particle-in-cell simulations of two-dimensional pair plasma.
High energy particles accelerated by the shock are found to drive magnetic
field evolution on a timescale $\ga 10^4$ plasma times.
Progressively stronger magnetic fields are generated on larger scales in a
growing region around the shock.
Shock-generated magnetic fields and accelerated particles carry $\gtrsim 1\%$
and $\gtrsim 10\%$ of the downstream energy flux, respectively. Our results
suggest limits on the magnetization of relativistic astrophysical flows.
\end{abstract}

\keywords{shock waves --- magnetic fields --- acceleration of particles --- gamma rays: bursts}

%\pacs{95.30.Qd, 52.35.Tc, 52.27.Ny, 98.70.Sa}

\maketitle

Due to the low plasma densities, shock waves observed in a wide range of
astronomical systems are collisionless, \ie mediated by collective plasma
instabilities rather than by binary particle collisions. Such shocks
play a central role in, for example, supernova remnants
\citep{Blandford_Eichler_87}, jets of radio galaxies
\citep{Maraschi_03}, \gama-ray bursts
\citep[GRBs,][]{Piran_2005}, pulsar wind nebulae
\citep[PWN,][]{kirk_2007}, and the formation of large-scale structure in
the Universe \citep{Loeb_Waxman_00}. It is widely accepted that particles
accelerated to high energy in such shocks generate the nonthermal
radiation observed in a wide range of astrophysical sources and
constitute the observed population of cosmic rays.

Despite intense research, collisionless shocks are still not understood
from first principles. In particular, there is no self-consistent theory
describing the acceleration of particles and the generation of magnetic
field fluctuations, which in turn scatter particles and mediate their acceleration. Much of
the research has focused on ``magnetized'' shocks, where the upstream magnetic
energy flux constitutes a significant fraction of the total energy flux.
Here we focus on ``unmagnetized'' shocks, where the upstream magnetic
energy flux is small. An extreme example of such shocks are the
relativistic GRB afterglow shocks, where the magnetic fraction of the
energy flux, $\epsilon_B$, increases from $\sim10^{-10}$ in the
upstream to $\epsilon_B\simeq 0.01-0.1$ in the downstream
\citep[][and references therein]{Waxman_06}. Under such conditions,
it is likely that the initial upstream magnetic field does not play a role in the
determination of the shock structure \citep[\eg][]{Gruzinov_2001}.

A near equipartition field, $\epsilon_B\simeq0.1$, may be produced by
electromagnetic (\eg Weibel) instabilities
\citep[\eg][]{Blandford_Eichler_87, Gruzinov_Waxman_99,Medvedev_Loeb_99}.
The coherence length of the generated field is expected to be comparable in this case to
the plasma skin-depth, $l_{sd}=c/\omega_p$, where $\omega_p$
is the plasma frequency and $c$ is the speed of light. The main challenge associated with the
downstream magnetic field is that while magnetic power on $\lsd$ scales rapidly
decays after the shock, observations imply that near equipartition fields must persist
over $10^{10}l_{sd}$ downstream \citep{Gruzinov_Waxman_99,Gruzinov_2001}.
This suggests that the magnetic field develops similarly large coherence lengths
\citep{Gruzinov_Waxman_99, Gruzinov_2001}. Such evolution could be driven by large
scale currents carried by particles accelerated in the shock, possibly leading
to a self-similar plasma configuration \citep{Katz_etal_07}. The mechanism
for the generation of large-scale magnetic fields in shocks remains unknown.

Particle-in-cell (PIC) simulations, in which the plasma is represented by
macroparticles and Maxwell's equations are solved on a grid, have been
extensively used in recent years to study shocks. Such studies have
quantified the generation of upstream current filaments by pinching
instabilities \citep[\eg][]{Silva_etal_2003, Frederiksen_etal_2004, Jaroschek_etal_2005,
Spitkovsky_2005, Spitkovsky_2008, Chang_etal_08},
and resolved the formation of shocks in
two- and three-dimensional (2D and 3D) pair plasma \citep{Spitkovsky_2005,
Kato_2007, Chang_etal_08} and in 2D ion-electron plasma
\citep{Spitkovsky_2008}. These simulations revealed rapid decay of magnetic
fields downstream \citep{Gruzinov_2001, Chang_etal_08}, leaving the
question of field survival over scales $\gg\lsd$ open and triggering
alternative suggestions for field generation
\citep[\eg][]{Goodman_2007,Milosavljevic_2007,Sironi_2007}.

In this \emph{Letter}, we report new PIC shock simulations performed on
unprecedentedly long length and time scales, $(L/l_{sd})^2 (T\omega_p) \simeq 4\times
 10^{10}$. These simulations show the growth of magnetic power on
progressively longer scales driven by the accelerated particles, and impose
lower limits on the efficiencies of particle acceleration and magnetization.
Our results suggest that even the most extensive simulations previously
reported \citep{Chang_etal_08} were too small to capture significant particle
acceleration and the resulting magnetic field evolution
\citep[see][for a discussion of PIC simulation results and limitations]{Katz_etal_07}.
Here we discuss only the main properties of shock evolution; the
particle acceleration mechanism is discussed separately \citep{Spitkovsky_2008_acc},
and a detailed analysis of the simulations is deferred to
a later publication.

\emph{Simulation set-up.} For simplicity, we focus here on strong,
relativistic shocks in unmagnetized pair plasma. In order to reach long
length and time scales, we resort to 2D, and comment below on expected
differences with respect to 3D shocks. We use the electromagnetic PIC code
TRISTAN-MP \citep{Spitkovsky_2005}, a parallel version of TRISTAN
\citep{Buneman_1993} heavily modified to minimize noise and numerical
instabilities. A rectangular simulation box is set up in the $x-y$ plane,
with periodic boundary conditions in the $y$-direction and a conducting
wall at $x_{wall}=0$. Cold, neutral plasma is continuously injected from
$x_{inj}=ct$ in the $-\hbx$ direction, where $t$ is the simulation time.
Reflection off the wall then results in a shock propagating along $+\hbx$.
All parameters are measured in the downstream frame, in which the wall is
at rest.

Typical simulation parameters are: injected bulk Lorentz factor
$\gamma_0=15$, thermal spread $\Delta\gamma_0=10^{-4}$, and $N_{ppc}=8$ particles
per species per cell, with spatial and temporal resolutions $\delta
x=l_{sd}/10$ and $\delta t=0.045\omega_p^{-1}$.
Here $\omega_p^2=4\pi
(n_{e^+}+n_{e^-}) q^2/\gamma_0 m$, where $m$ and $q$ are the particle
mass and charge, and $n$ is the upstream number density.
Our largest simulation has $\sim 2\times 10^{10}$
particles and $(L_x/\lsd)\times (L_y/\lsd) \times (T\omp) = 6300 \times
1024 \times 6300 \simeq 4\times 10^{10}$,
although smaller simulation boxes have been evolved for as long as
$12600\omp^{-1}$.
The results displayed below mostly refer to a simulation with $L_y=402 \lsd$,
evolved for $T=12600\omp^{-1}$.

\emph{Short time evolution}.
At early times, $t\lesssim 1000\omega_p^{-1}$, we recover shock formation
as reported previously \citep{Spitkovsky_2005, Chang_etal_08,
Spitkovsky_2008}: a transition layer of a few $10\lsd$ thickness
propagating upstream, in which the plasma isotropizes, thermalizes and
compresses. The simulated shock transition agrees to within a few percent
with (magnetic free) hydrodynamic jump conditions: a shock velocity $v_{sh}
= c(\Gamma_d-1)[(\gamma_0-1)/(\gamma_0+1)]^{1/2}$ and density compression
ratio $n_d/n_u =(\Gamma_d+\gamma_0^{-1})/(\Gamma_d-1)$. Here,
$\Gamma_d\simeq 3/2$ is the downstream adiabatic index, and upstream
pressure was neglected \citep{Spitkovsky_2008}.

Upstream, the interaction between the unshocked flow and a counterstream
running ahead of the shock leads to the formation of current filaments (in
both 2D and 3D) parallel to the flow, surrounded by near-equipartition
filamentary magnetic (in the fluid frame) structures. Behind the shock,
near equipartition magnetic clumps form and are advected with the
downstream flow in 2D. At early times (where 3D simulations are possible, $t\lesssim 10^3 \omp^{-1}$), good
agreement is found between these clumps and the 2D projection of extended
magnetic loops formed nearly perpendicular to the flow in 3D shocks.
When averaged along the transverse direction, $\epsilon_B \equiv
(B^2/8\pi)/[(\gamma_0-1)nmc^2]$ (where $B$ is the magnetic field amplitude)
peaks at $\sim 7\%$ near the shock transition layer and decays below
$0.1\%$ within $1000\lsd$ downstream \citep{Chang_etal_08}.

\emph{Long time evolution.} The above description does not include the
effects of high energy particles accelerated by the shock, negligible at
early times. Our present simulations are sufficiently large to reveal the
onset of particle acceleration and the slow evolution of shock properties
(evident on $\sim 1000\omp^{-1}$ timescales) driven by these energetic
particles. A small fraction of particles, accelerated to Lorentz factors
$\gamma_0\ll\gamma\ll \gamma_{max}$ by repeated scatterings near the
shock, gradually builds up a flat ($\gamma^2dn/d\gamma\sim \mbox{const.}$)
power-law energy tail downstream, already containing a fraction
$\epsilon_{acc}\gtrsim 10\%$ of the energy at
$t=10^4\omp^{-1}$ \citep{Spitkovsky_2008_acc}. Here we defined
$\epsilon_{acc}$ as the ratio between the energy density of particles with
$\gamma>5\gamma_0$ behind the shock, and the far upstream kinetic energy density,
such that for a thermal distribution of the particles in the downstream
$\epsilon_{acc}(\gamma_0\gg1)\simeq 0.3\%$.

\begin{figure}[h]
\centerline{\epsfxsize=10cm \epsfbox{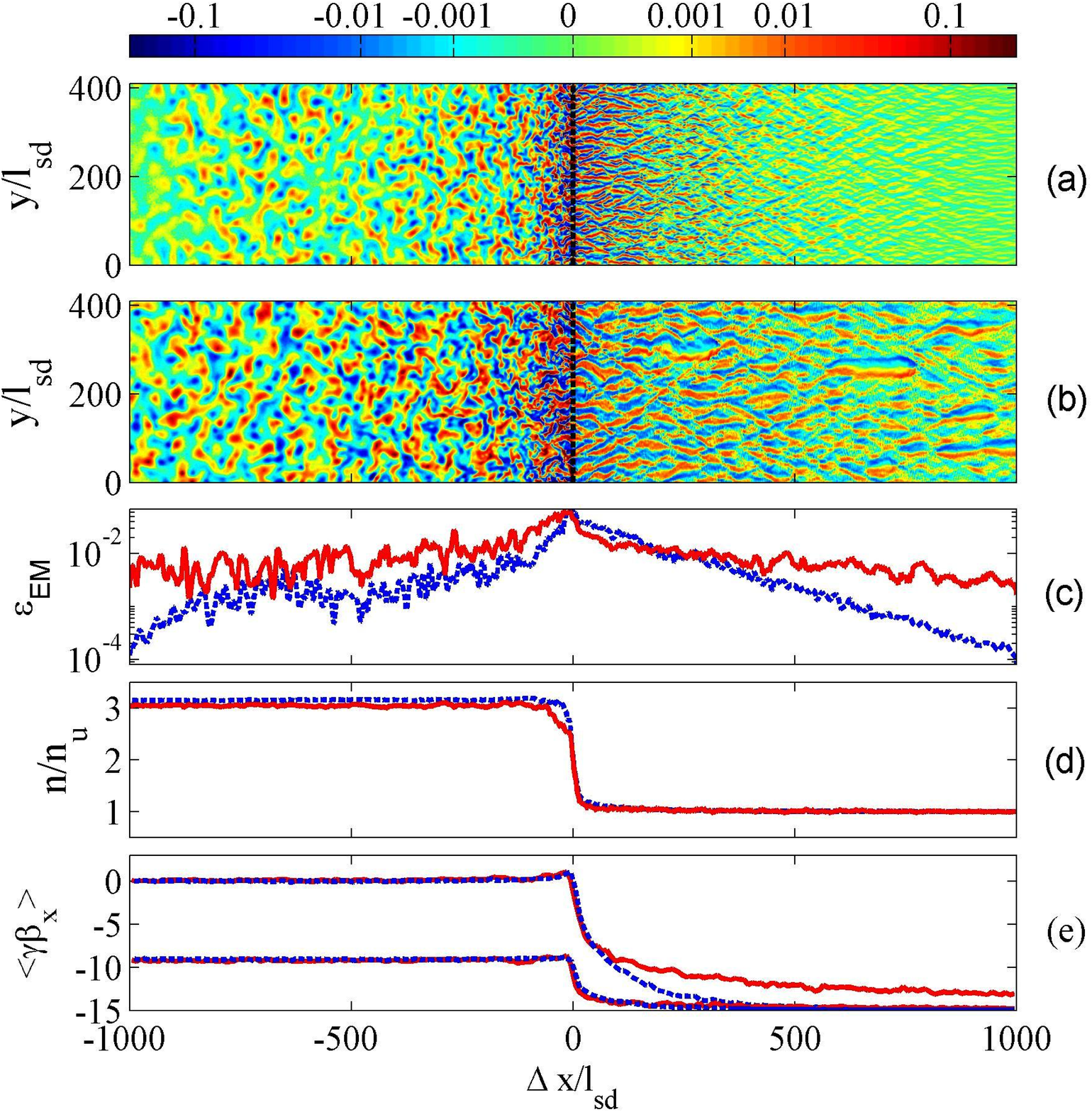} }
\caption{ Plasma evolution within $1000\lsd$ of the shock.
Normalized transverse magnetic field $\mbox{sign}(B)\epsilon_B$ (color
scale stretched in proportion to $\epsilon_B^{1/4}$ to highlight weak
features) is shown at (a) early ($t_1=2250\omp^{-1}$), and
(b) late ($t_2=11925\omp^{-1}$) times.
Here $\Delta x\equiv x-x_{sh}$ is the distance from the shock, with $x_{sh}$ (dashed)
defined as median density between far upstream and far downstream.
Also shown are the transverse averages (at $t_1$, dashed blue, and $t_2$, solid red)
of (c) electromagnetic energy normalized to the upstream kinetic energy,
$\epsilon_{EM} \equiv [(B^2+E^2)/8\pi]/[(\gamma_0-1)nmc^2]$
(with $E$ the electric field amplitude, included because  in the simulation
frame the induced $E\sim B$ upstream),
(d) density normalized to the far upstream, and (e) particle momentum
$\gamma\beta_x$ (with $\beta$ the velocity in $c$ units) in the
x-direction averaged over all particles (higher $\ave{\gamma \beta_x}$)
and over downstream-headed particles only.}
\label{fig1}
\end{figure}

The energetic particles running ahead of the shock significantly alter the
properties of the counterstream and the resulting current filamentation
and magnetization upstream. Figure \ref{fig1} shows the resulting spatial
distribution of magnetic fields at early vs. late times, as well as the
density and momentum profiles. It reveals an increasing magnetization
level, with fields generated on gradually larger scales and extending
farther away from the shock, both upstream and downstream. As a result,
the shock compression transition layer (defined, say, between $10\%$ and
$90\%$ of full shock compression) widens, $n$ and $\epsilon_B$ become more oscillatory
with distance behind the shock, the shock slightly accelerates (by
$\la 1\%$), and the final compression ratio slightly decreases (by $\la 4\%$).
Due to the substantial energy carried by the accelerated particles running
ahead of the shock, the average momentum is strongly modified far
upstream, $\Delta x \equiv x-x_{sh}\gtrsim 1000 \lsd$, although the
incoming flow slows down considerably only at $\Delta x\lesssim 100\lsd$
(figure 1e).

\begin{figure}[h]
\centerline{\epsfxsize=10cm \epsfbox{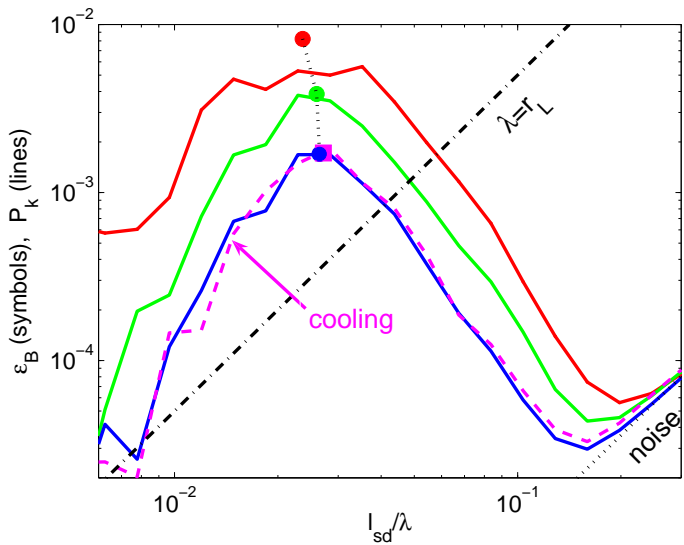}}
\caption{ Magnetic 2D power spectrum in a sample downstream region defined
by $-1000<\Delta x/l_{sd}<-200$. The spectrum is shown to gradually grow and
possibly flatten with time [solid curves for $t\omp\simeq 1900$ (blue), $4600$
(green), and $12600$ (red)].
The integrated energy fraction $\epsilon_B$ and average scale
$\ave{\lambda}\equiv 2\pi/\exp[\epsilon_B^{-1}\int \log(k)\,P_k\,d\log k]$ of
the magnetic field (filled circles) grow correspondingly.
Suppressing particle acceleration (cooling all particles above
$\gamma_{cool}=80$, dashed line and square for $t\omega_p=5750$) stops
the magnetic evolution. Also shown are estimated shot-noise power (dotted)
and a $\lambda=r_L\simeq(2P_k)^{-1/2}\lsd$ curve (dash
dotted) with $r_L$ the Larmor radius of $\gamma=\gamma_0$ particles,
roughly separating magnetized (large scale) and
non-magnetized (small scale) bulk plasma regimes. }\label{fig2}
\end{figure}

Figure \ref{fig2} quantifies the evolution of downstream magnetization. It
shows the power spectrum $P_k$ defined by $\epsilon_B=\int P_k \, d\log k$ (curves), the average
coherence length $\ave{\lambda}$ and the energy fraction $\epsilon_B$
(filled circles) of magnetic fields in an $l_x=800\lsd$ long region trailing
behind the shock. In order to avoid shot noise
contamination\footnote{Noise with power inversely proportional to $N_{ppc}$,
filtered on small scales, and tested not to distort our results.},
$\epsilon_B$ is measured only for coherence lengths
$\lambda=2\pi/k>10\lsd$. During $10^3\la t\omp\la 10^4$, $\epsilon_B$
in this region grows by a factor of $\sim 4$ and reaches $\sim 1\%$. As
illustrated in Figure \ref{fig1}, the typical size of upstream filaments
increases substantially in time, with typical thickness $(10-15)\lsd$ at
$t=10^{3}\omp^{-1}$ growing by a factor of $3-4$ by
$t=10^4\omp^{-1}$. Figure \ref{fig2} shows a modest effect downstream,
where the average magnetic coherence length grows by $\sim 10\%$ during
this epoch.

In order to test the role played by high energy particles in the evolution
of the shock, we have performed a suite of simulations with artificially
suppressed particle acceleration. In these runs, we introduced cooling, where
particles with $\gamma>\gamma_{cool}$ lose a random fraction of their
nonthermal energy, with various choices of $\gamma_{cool}$.
Cooling is found to significantly slow down or completely stop shock
evolution, leading to a fixed magnetization level and a steady-state
magnetic power spectrum. Higher values of $\gamma_{cool}$ are found to
produce larger $\epsilon_B$ and $\ave{\lambda}$. For example, a steady
state configuration with $\epsilon_B=0.2\%$ (at $-1000<\Delta x/\lsd<-200$), obtained for
$\gamma_{cool}=80$ ($\gamma_0=15$), is shown as a dashed line in Figure \ref{fig2}.

\begin{figure}[h]
\centerline{\epsfxsize=10cm \epsfbox{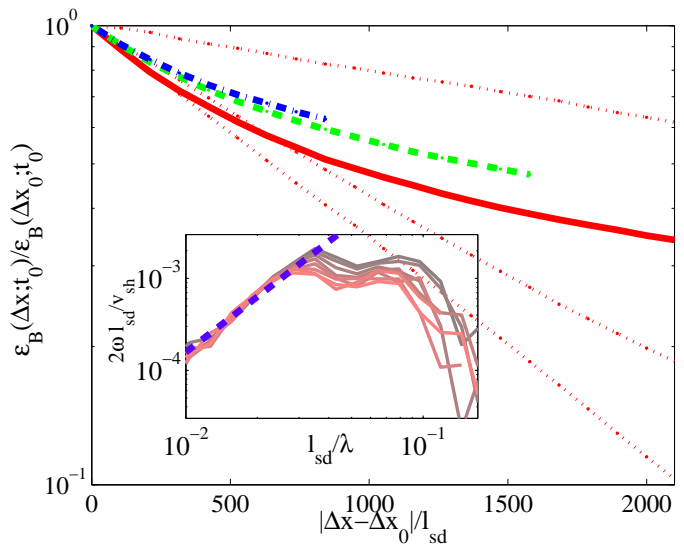}} \caption{ Decay of magnetic
energy in three different slices advected downstream at progressively later times.
Each slice is
$l_x=1000\lsd$ long, located with its upstream edge at ${\Delta}x_0 =
-500\lsd$ at $t_0\omp=3600$ (solid red), $5175$ (dashed green) and 6750
(dash-dotted blue); all curves cutoff when the simulation terminates at
$t\omp=8550$.
The decay rate is seen to slow down as the shock evolves.
Power on scales $\lambda\gtrsim 40\lsd$ decays nearly exponentially in
$|\Delta x|$, shown (dotted red) for $t_0\omp=3600$ and $\lambda/\lsd=40$,
$50$ and $67$ (fast to slow decay).
In \cite{Chang_etal_08}, magnetic decay was studied in a fixed downstream
frame region (\eg solid line), not taking into account magnetic field
evolution due to accelerated particles. \newline
Inset: Spatial decay rate of the magnetic power spectrum $(2/v_{sh})\omega(k) =
P_k^{-1}(-dP_k/d{\Delta}x)$ in a slice (with $l_x=1400\lsd$ and
$t_0=4500\omp^{-1}$) advected downstream at $(t-t_0)\omp=0, 225, 450,
\ldots 1800$ (solid, dark to light curves). At large scales, the decay
rate is well fitted by $\omega\propto k^2$ (dashed), whereas at small
scales $\omega$ decreases with time/distance from the shock. }
\label{fig3}
\end{figure}

The decay rate of magnetic fields advected downstream slows down as the
shock evolves, as illustrated in Figure \ref{fig3}. This is partly attributed
to the increased fraction of power deposited in large scale fields, which
are expected to survive farther downstream. Indeed, the inset of Figure
\ref{fig3} shows that for $\lambda\gtrsim 40\lsd$, magnetic evolution is
well fitted by exponential decay, $P_k \propto e^{-\omega t}$, with
$\omega \propto k^2$. Such behavior is expected, for example, in MHD
magnetic diffusion with scale-independent resistivity (note that Figure
\ref{fig2} suggests that the bulk plasma is indeed magnetized on these
scales). Figure \ref{fig3} thus indicates, for example, that power on
$\lambda=30\lsd$ ($100\lsd)$ scales survives more
than $10^3\lsd$ ($10^4\lsd$) downstream.

\emph{Convergence.}
Convergence tests were performed with respect to all simulation parameters
(around their values given above), with no qualitative changes to the
results.
Figures \ref{fig1} and \ref{fig2} indicate that the
simulation box used is sufficiently large to avoid significant boundary
effects. However, as $\sim 40\%$ ($5\%$) of the magnetic power is already
deposited in $\lambda>50\lsd$ ($\lambda>100\lsd$) scales at $t\simeq
10^4\omp^{-1}$, increasingly larger simulation boxes, in both longitudinal
and transverse dimensions, will be required in order to properly resolve the
shock and the growing coherent structures.

\emph{Conclusions.} Our analysis shows that collisionless shock
configurations simulated previously may represent steady state configurations
only as long as
particle acceleration remains insignificant. We find that a population
of energetic particles is accelerated and drives
the generation of progressively stronger fields on gradually larger scales.
Our simulations do not reach a steady state; rather, an increasing
fraction of shock energy is transferred to energetic particles and
magnetic fields throughout the simulation time domain.

Once stochastic acceleration and magnetization ensue, they are unlikely to
diminish to a lower energy steady state. Hence, our results suggest
lower limits to the efficiencies of magnetization and particle acceleration,
$\epsilon_B\gtrsim 1\%$ at distances $|\Delta x|<D=1000\lsd$ downstream of the
shock, and $\epsilon_{acc}\gtrsim 10\%$ with no significant cooling identified
downstream. We find no evidence for the saturation of $\epsilon_{acc}$,
$\epsilon_B$, $D$ or $\gamma_{max}$, although the high energy particles
downstream are already sub-equipartition at $t\simeq 10^4\omp^{-1}$.

Although our results are obtained for 2D pair plasma, we expect qualitatively
similar shock evolution in 3D shocks and in electron-ion plasma.
While the nature of upstream current filaments and downstream magnetic
loops/clumps may depend on dimensionality, 2D and 3D simulations are
in good agreement at early times (Spitkovsky \& Arons, in preparation).
Also, shocks in ion-electron plasma were found to be similar to pair
plasma shocks at early-time 2D simulations due to efficient electron
heating \citep{Spitkovsky_2008}. Some level of stochastic particle
acceleration is inevitable in all cases, but the generalization of our
results to 3D or to ion-electron plasma is yet to be tested at late times
and in the presence of a high energy particle tail.

The major role played by high energy particles in shock evolution, their
flat spectrum and the apparently flattening magnetic power spectrum are
trends consistent with a self-similar plasma configuration
\citep{Katz_etal_07}, although the simulated downstream scale growth is
more modest than the $\lambda\propto D$ self-similar scaling. At this
stage, the simulations are not yet sufficiently advanced to validate or
rule out self-similarity.

In summary, we have shown that collisionless shocks in 2D pair plasma
evolve on long, $\gtrsim 10^3 \omp^{-1}$ timescales, such that the
acceleration efficiency, magnetization level, and coherence length scale
all increase in time. These trends and the above lower limits on
$\epsilon_{acc}$ and $\epsilon_B$ indicate that a shock propagating into a cold,
homogeneous plasma with $B=0$ remains a viable model for astronomical
shocks, with no need for additional assumptions about magnetic turbulence generation
\citep[\eg][]{Goodman_2007, Milosavljevic_2007, Sironi_2007}.
Our results confirm that particle acceleration
and magnetization are intimately related, with high energy particles playing a
major role in generating the magnetic fields which in turn scatter and
accelerate the particles.

\acknowledgements We thank J. Arons and P. Goldreich for helpful
discussions. U.K. is a Friends of the Institute for Advanced Study member, and
is supported by NSF grant PHY-0503584. The research of B. K. \& E.
W. is partially supported by AEC, ISF \& Minerva grants. A.S. acknowledges
the use of computational resources at TIGRESS computing center at Princeton
University and the support from Alfred P. Sloan Foundation fellowship.

\bibliographystyle{apj}

\end{document}